\documentclass{PoS}

\usepackage{amsmath}
\usepackage{amssymb}
\usepackage{epsfig}

\PoS{PoS(LAT2005)308}
\title{Anatomy of string breaking in QCD}
\ShortTitle{Anatomy of string breaking}
\author{Zdravko Prkacin${}^1$, \speaker{Gunnar S.~Bali}${}^{,2}$, 
Thomas D\"ussel${}^3$,
Thomas Lippert${}^{1,3}$,
Hartmut Neff${}^4$ and Klaus Schilling${}^2$\\
${}^1$Fachbereich Physik, Bergische Universit\"at Wuppertal -
D-42097 Wuppertal, Germany\\
${}^2$Dept.\ of Physics and Astronomy, The University
of Glasgow - Glasgow G12 8QQ, UK\\
${}^3$Zentralinst.\ f.\ Angewandte Mathematik,
Forschungszentrum J\"ulich - D-52425 J\"ulich, Germany\\
${}^4$CCS,
Chemistry Dept., Univ.\ College London -
20 Gordon Street, London WC1H 0AJ, UK\\
E-mail:
\email{prkacin@theorie.physik.uni-wuppertal.de},
\email{g.bali@physics.gla.ac.uk},
\email{th.duessel@fz-juelich.de},
\email{th.lippert@fz-juelich.de},
\email{uccahne@ucl.ac.uk},
\email{schillin@theorie.physik.uni-wuppertal.de}
}

\abstract{We investigate the string breaking mechanism in $n_f=2$ QCD.
We discuss the lattice techniques used and present results on
energy levels and mixing angle of the static $B\overline{B}|\overline{Q}Q$
two-state system. The string breaking is visualized,
by means of an animation of the action density distribution as
a function of the static colour source-antisource separation.}

\FullConference{XXIII International Symposium on Lattice Field Theory
(Lattice 2005)\\Dublin, Ireland\\July, 25-30 2005}

\begin{document}
\newcommand{\www}{\mbox{
\begin{minipage}{16pt}\includegraphics[width=16pt]{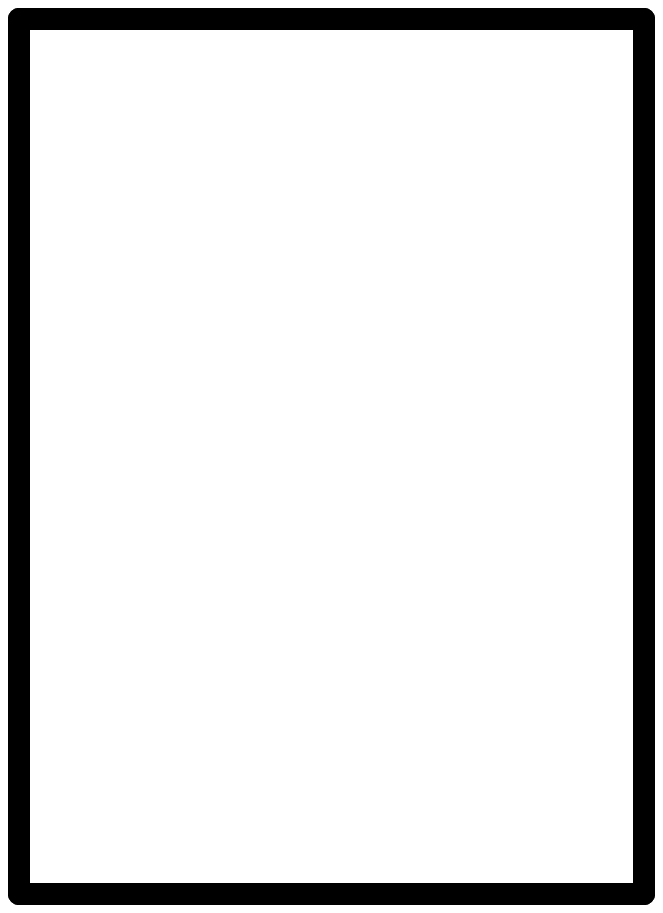}\end{minipage}}}
\newcommand{\wbbc}{\mbox{
\begin{minipage}{16pt}\includegraphics[width=16pt]{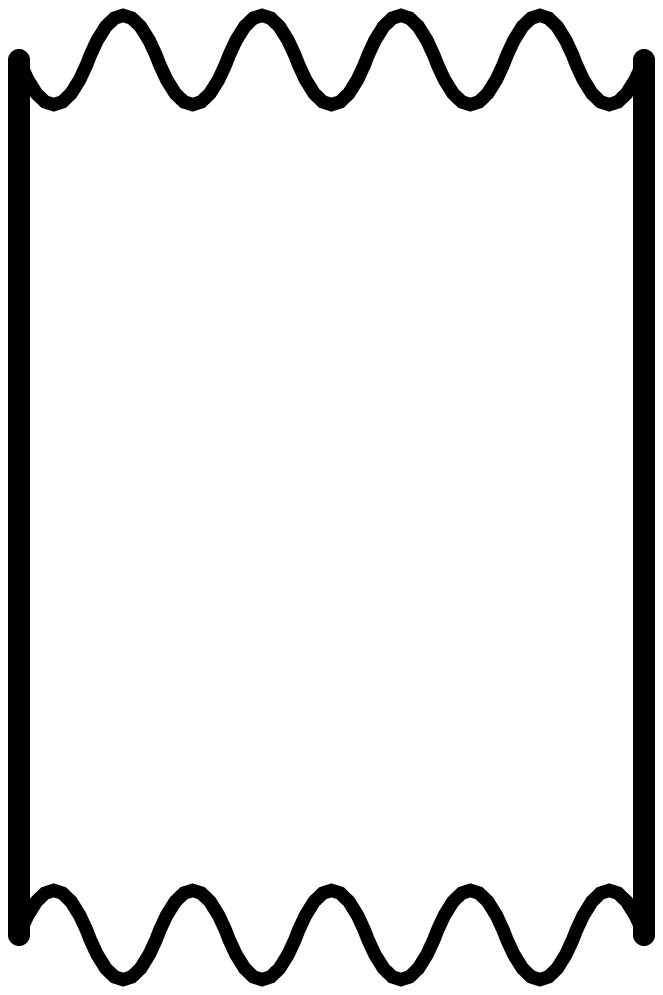}\end{minipage}}}
\newcommand{\wbbd}{\mbox{
\begin{minipage}{16pt}\includegraphics[width=16pt]{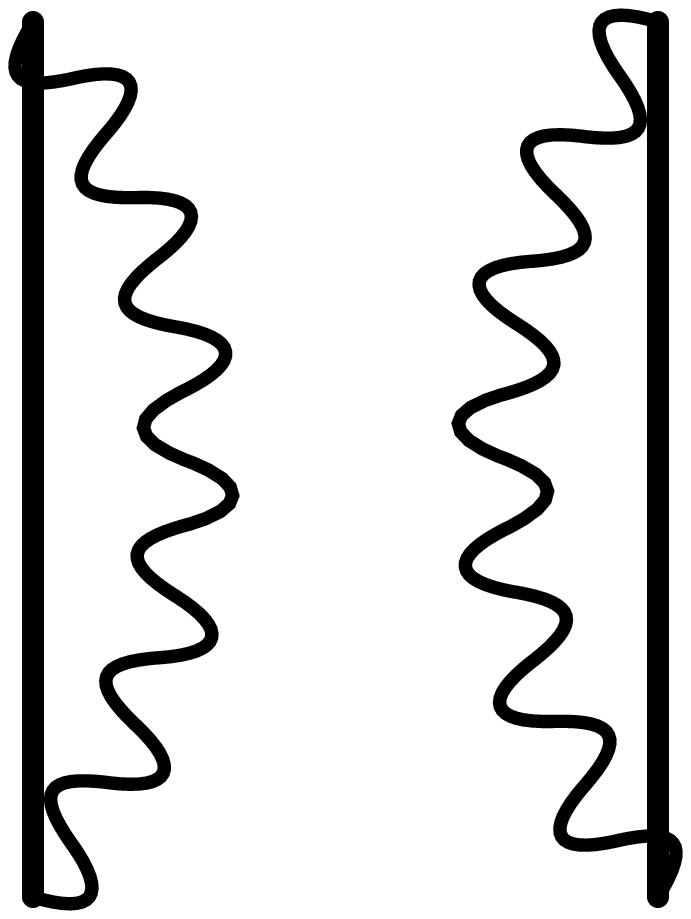}\end{minipage}}}
\newcommand{\wwb}{\mbox{
\begin{minipage}{16pt}\includegraphics[width=16pt]{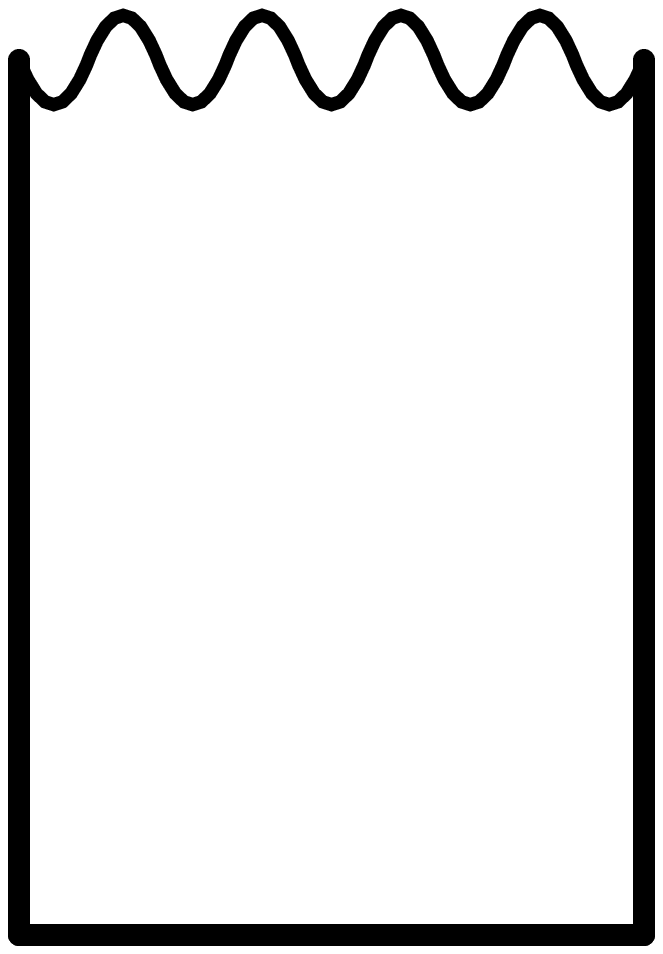}\end{minipage}}}
\newcommand{\wbw}{\mbox{
\begin{minipage}{16pt}\includegraphics[width=16pt]{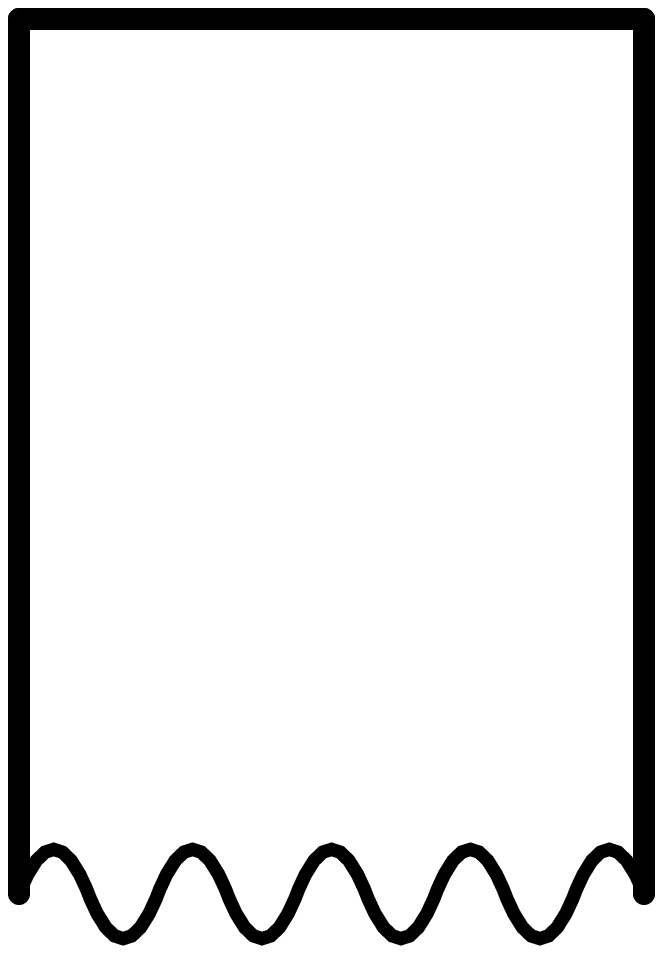}\end{minipage}}}
\section{Introduction}
The breaking of the colour-electric string between two
static sources is a prime example of a
strong decay in QCD~\cite{Michael:2005kw}. Recently, we
reported on an investigation of this two state
system~\cite{Bali:2004pb,Bali:2005fu},
with a wave function
$|Q\rangle$ created by a $\overline{Q}Q$ operator
and a wave function $|B\rangle$ created by a four-quark $B\overline{B}$
operator, where $B=\overline{Q}q$. $Q$ denotes a static
source and $q$ is a light quark.
We determined the energy levels
$E_1(r)-2m_B$ and $E_2(r)-2m_B$ of the
two physical eigenstates $|1\rangle$ and $|2\rangle$ which
we decomposed into the components,
\begin{eqnarray}
\label{eq:1}
|1\rangle&=&\cos{\theta}\,|Q\rangle+\sin{\theta}\,|B\rangle\\
|2\rangle&=&-\sin{\theta}\,|Q\rangle+\cos{\theta}\,|B\rangle.
\label{eq:2}
\end{eqnarray}
We characterize string breaking
by the distance scale $r_c$ at which $\Delta E = E_2-E_1$
is minimized and by the
energy gap $\Delta E_c=\Delta E(r_c)$. While these energy levels
and $r_c$ are {\em first principles}
QCD predictions, the mixing angle $\theta$ is (slightly)
model dependent: within each (Fock) sector there are further
radial and gluonic excitations and we truncated the basis
after the four quark operator.

We use $n_f=2$ Wilson fermions at a quark mass
slightly smaller
than the physical strange quark and a lattice spacing
$a=0.166(2)\,r_0\approx 0.083(1)$~fm and
find~\cite{Bali:2005fu}, $r_c=2.5(3)\,r_0\approx 1.25(1)\,\mbox{fm}$
and $\Delta E_c\approx 51(3)$~MeV, where the errors do
not reflect the phenomenological uncertainty of assigning a physical
scale to $r_0\approx 0.5$~fm.
Using data on the $\overline{Q}Q$ potential and the
static-light meson mass $m_B$, obtained at different quark masses,
we determine the real world estimate, $r_c= 1.13(10)(10)$~fm,
where the errors reflect all systematics. An extrapolation
of $\Delta E_c$ however is impossible,
without additional
simulations at lighter quark masses.

These results became possible by combining a variety of improvement
techniques: the necessary all-to-all light quark propagators were
calculated from the lowest eigenmodes of the
Wilson-Dirac operator, multiplied by $\gamma_5$, after a
variance reduced stochastic estimator correction step. The signal
was improved by employing a fat link static
action. Many off-axis distances were implemented to allow
for a fine spatial resolution of the string breaking region.
Last but not least great care was taken to optimize
the overlap with the ground states, within
the $|Q\rangle$ and $|B\rangle$ sectors,
using combinations of three dimensional APE and Wuppertal smearing.
For details see Ref.~\cite{Bali:2005fu}.
We were able to achieve
values of 
$|a_Q|^2=|\langle \Psi_Q|Q\rangle|^2=0.62(2)$ and
$|a_B|^2=|\langle \Psi_B|B\rangle|^2=0.96(1)$ at
$r\approx r_c$ for the overlaps of
our test wave functions $|\Psi_X\rangle$
with the respective states
on the right hand sides of Eqs.~(\ref{eq:1}) and (\ref{eq:2}).
The almost optimal value of $|a_B|\leq 1$ was essential
to allow $E_1$, $E_2$ and $\theta$ to be fitted from
correlation matrix data,
\begin{equation}
C(t)=\left(\begin{array}{cc}C_{QQ}(t)&C_{QB}(t)\\
C_{BQ}(t)&C_{BB}(t)\end{array}\right)
=\left(\begin{array}{rl}
\quad \www&\quad \sqrt{n_f}\wwb\\&\\
\sqrt{n_f}\wbw&\quad -n_f\wbbc+\wbbd\end{array}\right),\label{eq:sb}
\end{equation}
obtained at moderate
Euclidean times:
$t\geq 2a$ for $C_{BB}$ and
$t\geq 4a$ for the remaining 2 matrix elements $C_{QQ}$ and
$C_{QB}=C_{BQ}$ at $r\approx r_c$.

In order to obtain dynamical information on the string breaking
mechanism,
we now study the spatial
energy and action density distributions within
the two state system. For instance one can then ask
questions about the localisation of the light $q\bar{q}$ pair
that is created when $r$ is increased beyond $r_c$.
The energy density will decrease fastest
in those places where $q\bar{q}$ creation is most likely.
Perturbation theory suggests that
light pair creation close to one of the static sources
is favoured by the Coulomb energy gain while aesthetic arguments
might suggest a symmetric situation with $q\bar{q}$
dominantly being created near the centre. 

Another motivation for a more detailed study
is our wish to relate the
static limit results to strong decay rates of
quarkonia. In the non-relativistic limit of heavy quarks,
potential ``models'' provide us with the natural
framework for such studies. In fact at short distances,
$r\ll 1/\Lambda$, potential ``models'' can be derived as
an effective field theory, potential NRQCD,
from QCD~\cite{Brambilla:2004jw}. One can easily add a
$B\overline{B}$ sector, as well as transition terms between the
two sectors, to the $\overline{Q}Q$ pNRQCD Lagrangian. Strong
decays would then be a straightforward non-perturbative
generalisation
of the standard multipole treatment of radiative transitions
in QED. Unfortunately, transitions such as
$\Upsilon(4S)\rightarrow B \overline{B}$ can hardly be classed
as ``short distance'' physics. So, some modelling is required.
The natural starting point again is a two channel
potential model which might still have some validity
beyond the short distance regime.
The transition rate would then be given by phase space
times an overlap integral
between the $\Upsilon(4S)$ wave function and the $B\overline{B}$
continuum, where some additional input is required
to constrain the sandwiched interaction term (for details
see e.g.\ Ref.~\cite{Drummond:1998eh}).

\section{The Method}
We follow Ref.~\cite{Bali:1994de} and
define action and energy density distributions,
\begin{eqnarray}\label{eq:sig}
\sigma_n({\mathbf x})=\frac12\left[{\mathcal E}_n({\mathbf x})+
{\mathcal B}_n({\mathbf x})\right],\\
\epsilon_n({\mathbf x})=\frac12\left[{\mathcal E}_n({\mathbf x})-{\mathcal B}_n({\mathbf x})\right],
\label{eq:eps}
\end{eqnarray}
where
\begin{equation}
{\mathcal A}_n({\mathbf x})=\langle n|A^2({\mathbf x})|n\rangle
-\langle A^2\rangle=\lim_{t\rightarrow\infty}\frac
{\langle \Phi_n(t)|A^2({\mathbf x,t/2})|\Phi_n(0)\rangle}
{\langle \Phi_n(t)|\Phi_n(0)\rangle}-\langle A^2\rangle
\label{eq:cal}
\end{equation}
We have suppressed the distance $r$ from the above formulae
and $n=1$ denotes the ground state (dominantly $\overline{Q}Q$
at $r<r_c$) and $n=2$ the excitation (dominantly $B\overline{B}$
at $r<r_c$).
Electric and magnetic fields are calculated from the
plaquette,
\begin{eqnarray}
E^2\left(x+\frac12a\hat{\mathbf 4}\right)=\frac{2\beta}{a^4} \frac{1}{2}\sum_{i=1}^3
\left[\overline{U}_{x,i4}+\overline{U}_{x-a\hat{\boldsymbol{\imath}},i4}\right],\\
B^2(x)=\frac{2\beta}{a^4} \frac{1}{4}\sum_{i=1}^3
\left[\overline{U}_{x,ij}(x)+\overline{U}_{x-a\hat{\boldsymbol{\imath}},ij}
+\overline{U}_{x-a\hat{\boldsymbol{\jmath}},ij}
+\overline{U}_{x-a\hat{\boldsymbol{\imath}}-a\hat{\boldsymbol{\jmath}},ij}
\right],
\end{eqnarray}
where $j=\mbox{mod}(i,3)+1$ and,
\begin{equation}
\label{eq:nn}
\overline{U}_{x,\mu\nu}=\frac{z_0}{3}\mbox{tr}\,\left(\overline{U}_{x,\mu}
\overline{U}_{x+a\hat{\boldsymbol{\mu}},\nu}
\overline{U}^{\dagger}_{x+a\hat{\boldsymbol{\nu}},\mu}
\overline{U}^{\dagger}_{x+a\hat{\boldsymbol{\mu}}+
a\hat{\boldsymbol{\nu}},\nu}\right).
\end{equation}
We implemented two different operators with the same
continuum limits: in one case we identified
$\overline{U}_{x,\mu}$ with the link
$U_{x,\mu}$ connecting $x$ with $x+a\hat{\boldsymbol{\mu}}$.
In addition we used smeared operators,
\begin{equation}
\overline{U}_{x,\mu}=
P_{SU(3)}\left(\gamma\,U_{x,\mu}+\sum_{|\nu|\neq \mu}
U_{x,\nu}U_{x+a\hat{\boldsymbol{\nu}},\mu}
U^{\dagger}_{x+a\hat{\boldsymbol{\mu}},\nu}\right),
\end{equation}
where $\gamma = 0.4$ and the sum is over all six staples, in the three
forward and three backward directions. The $\gamma$-value was tuned to maximize
the average plaquette, calculated from the smeared links.
$P_{SU(3)}$ is a projection operator
into the $SU(3)$ group. For the un-smeared plaquette
$z_0=1$ in Eq.~(\ref{eq:nn}) while for smeared plaquettes
$z_0=1+O(\alpha_s)$ is adjusted such that the vacuum expectation value
of the average plaquette remains unchanged.

The plaquette smearing enhances the signal/noise
ratio. Due to this smearing and the fat link static
action used, the peaks of the distributions around the source
positions (that will diverge in the continuum limit)
are less singular than in previous studies of
$SU(2)$ gauge theory at similar lattice
spacings~\cite{Bali:1994de}.
In the continuum limit the results from smeared
and un-smeared plaquette probes will coincide, away from
these self energy peaks. The draw back of plaquette smearing
is that exact reflection
positivity is violated. However, our wave functions are
sufficiently optimized to compensate for this.

We insert the $E^2({\mathbf x},t)$ and $B^2({\mathbf x},t)$ operators
at position $t/2$ into the correlation matrix $C(t)$, Eq.~(\ref{eq:sb}),
see Ref.~\cite{Bali:2005fu}. For
even|odd $t/a$-values
we average $E^2$|$B^2$ over the two adjacent time slices, respectively.
Using the fitted ground state overlap ratio $a_Q/a_B$ 
and the mixing angle $\theta$ as inputs, we
calculate the action and energy density
distributions Eqs.~(\ref{eq:sig}) and (\ref{eq:eps})
in the limit of large $t$ via Eq.~(\ref{eq:cal})
from the measured matrix elements. The distributions agree within errors
within the time range $3a\leq t \leq 6a$. The results presented here
are all based on our $t=4a$ analysis.

\section{Results}
\begin{figure}[th]
\begin{minipage}{.5\textwidth}
\epsfig{file=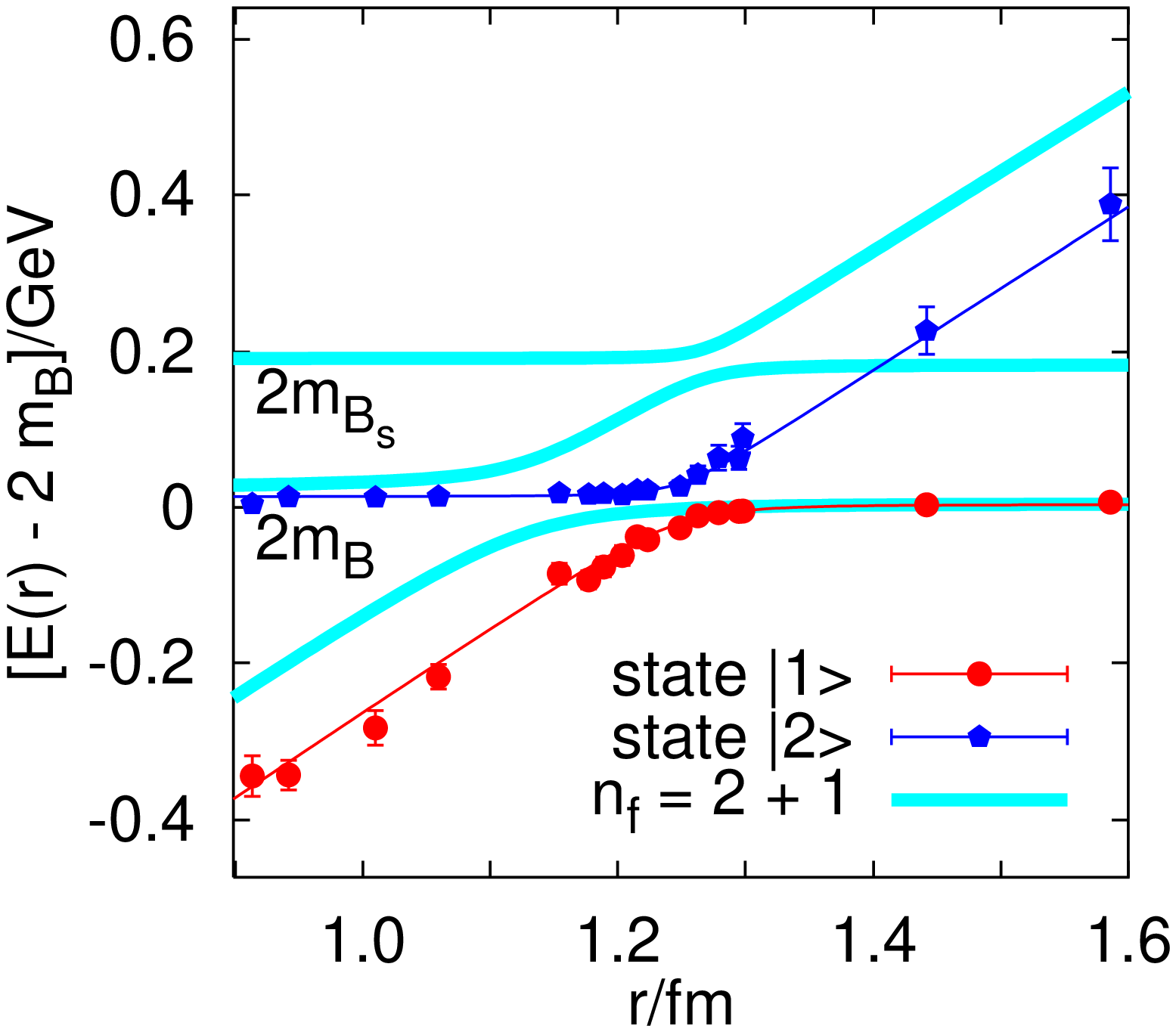,width=0.9\textwidth}
\end{minipage}
\begin{minipage}{.49\textwidth}
\epsfig{file=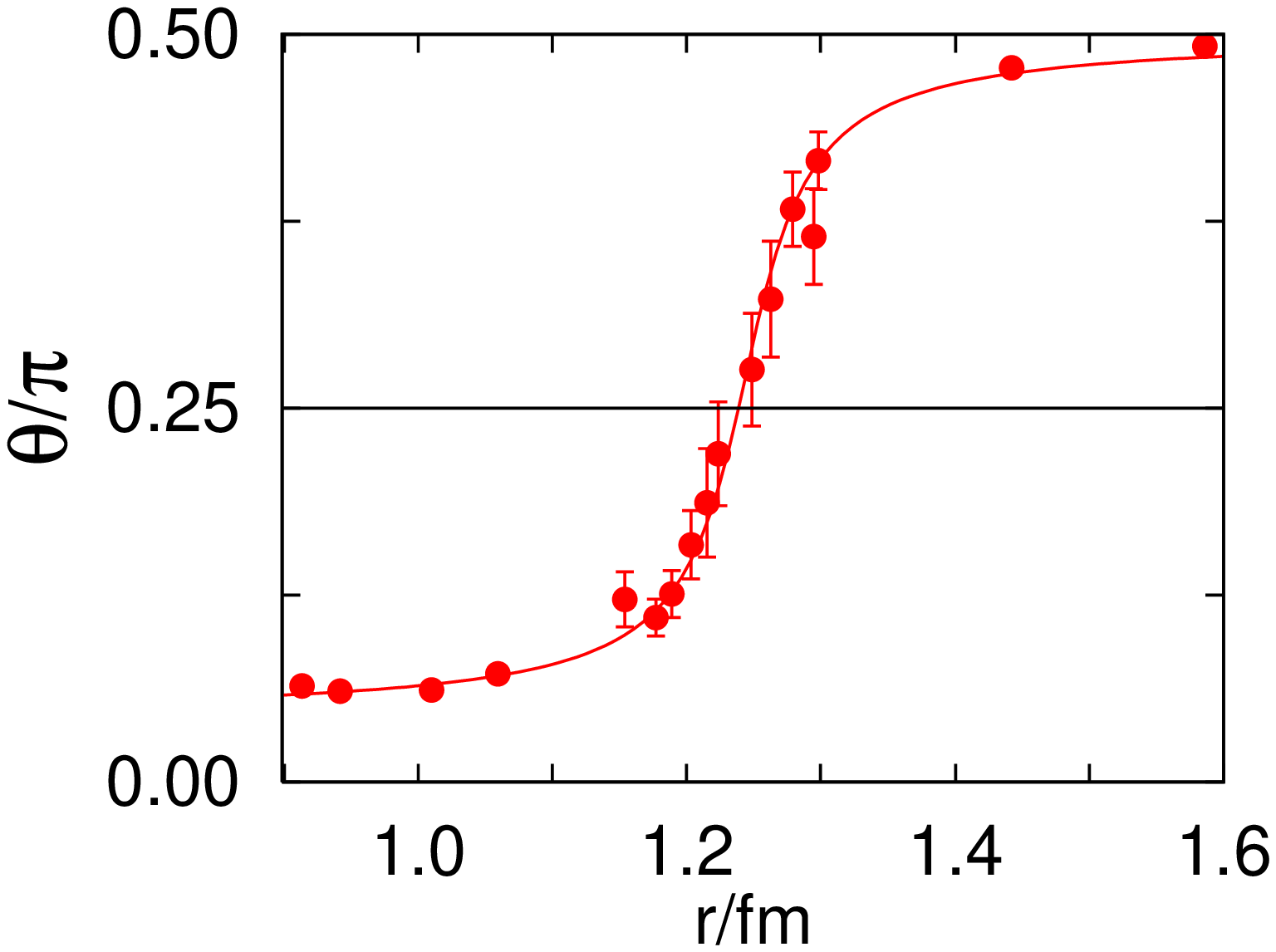,width=0.95\textwidth}
\end{minipage}
\caption{The energy levels and the mixing angle $\theta$
in physical units for $n_f=2$. The bands in the left figure
reflect the expected $n_f=2+1$ scenario.}
\label{fig:pot}
\end{figure}

To set the stage,
we display the main results of
Ref.~\cite{Bali:2005fu} in Figure~\ref{fig:pot}.
In the left figure we also speculate about the
scenario in the real world with possible decays into
$B\overline{B}$ as well as into $B_s\overline{B}_s$.
As discussed above, for our parameter
settings and $n_f=2$
string breaking occurs at a distance $r_c\approx 1.25$~fm.
In the right figure we show the mixing angle as a function of the
distance. The $B\overline{B}$ content of the ground state is given
by $\sin\theta$. Within our statistical errors $\theta$ reaches
$\pi/4$ at $r=r_c$. Remarkably, there is a significant four quark
component in the ground state at $r<r_c$ while for $r>r_c$ 
the limit $\theta\rightarrow\pi/2$ is rapidly approached.

\begin{figure}[th]
\centerline{\epsfig{file=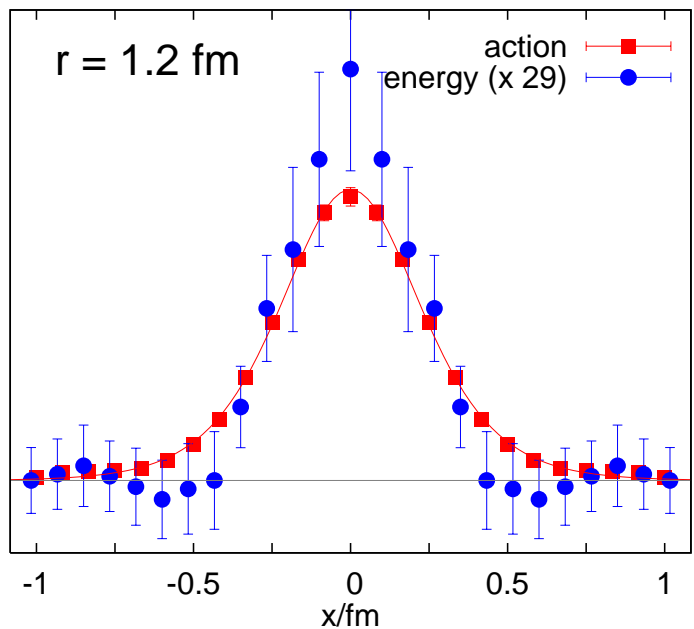,width=0.6\textwidth}}
\caption {The transverse profile in the centre of the flux tube.}
\label{fig:transverse}
\end{figure}
The quality of our density distribution data is depicted in
Figure~\ref{fig:transverse} for the ground state at
a separation slightly smaller than $r_c$, as
a function of the transverse distance $x$
from the $\overline{Q}Q$ axis. 
Due to cancellations between the magnetic and
electric components the energy density is much smaller than the action
density: for the comparison we
have multiplied the energy density data by the arbitrary factor of 29.
Note that the ratio $\sigma/\epsilon$ will diverge like $-\ln a\Lambda$
in the continuum limit. The differences between the shapes
of the energy and action density distributions are not statistically
significant. Here we only visualize the more precise action density results.

We employ several off-axis separations. Assuming
rotational symmetry about the interquark axis, each
point is labelled by two coordinates.
$x$ denotes the distance from the $\overline{Q}Q$ axis
and $y$ denotes the longitudinal distance from the centre point.
We define an interpolating rectangular
grid with perpendicular lattice spacing $a$ and the longitudinal spacing
slightly scaled, such that the static sources always lie on
integer grid coordinates. We then assign a quadratically interpolated value
to each grid point ${\mathbf z}$, obtained from points
in the neighbourhood,
$|{\mathbf z}-(x,y)|\leq \epsilon=a$.
On the axis the data points are more sparse and we relax the condition
to $\epsilon=\sqrt{3}a$ while for the singular peaks we maintain the
un-interpolated values.

\begin{figure}[th]
\begin{minipage}{.49\textwidth}
\href{http://pos.sissa.it//archive/conferences/020/308/LAT2005_308_a1.mpg}{\epsfig{file=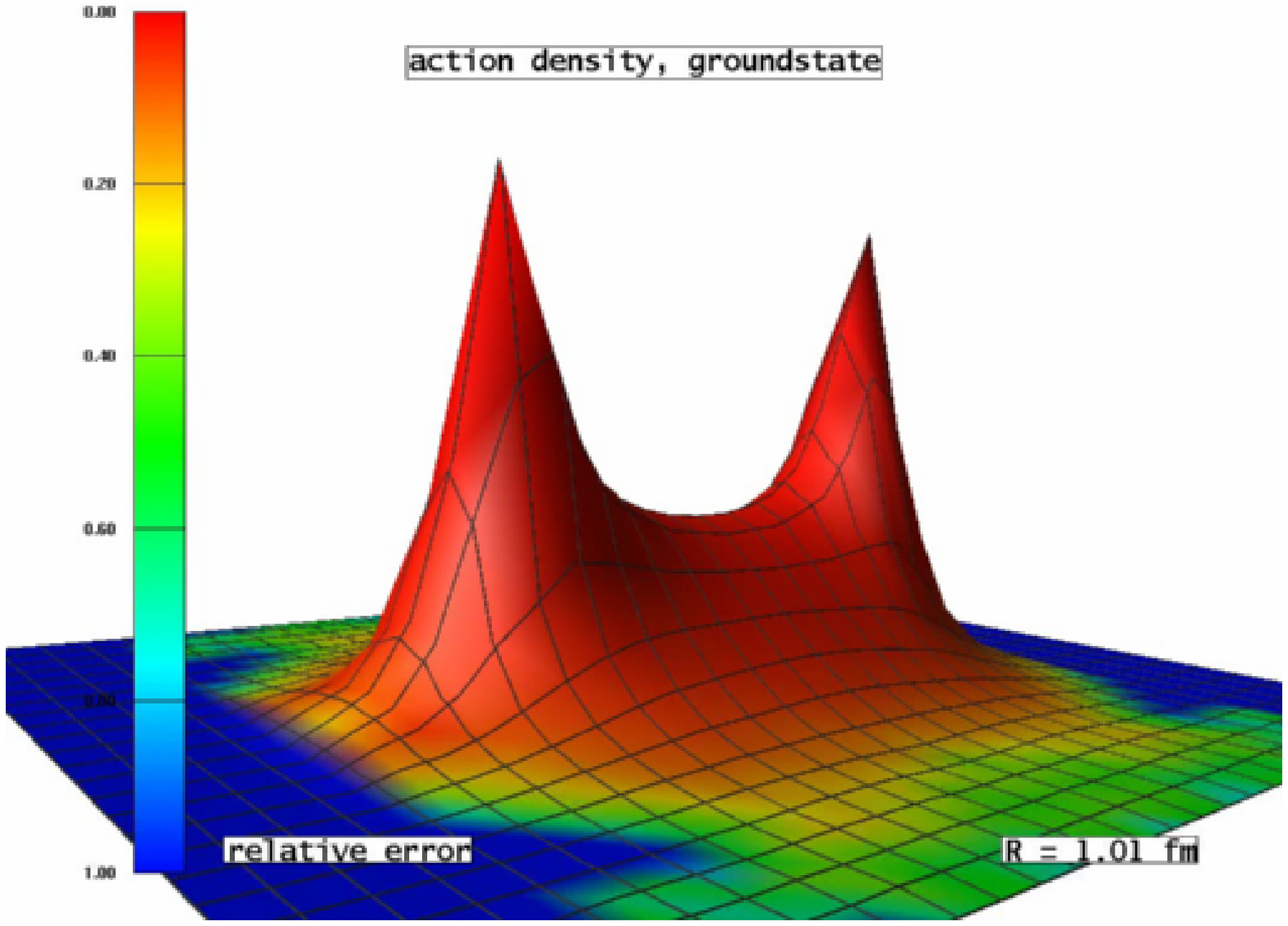,width=0.95\textwidth}}
\end{minipage}
\begin{minipage}{.49\textwidth}
\href{http://pos.sissa.it//archive/conferences/020/308/LAT2005_308_a2.mpg}{\epsfig{file=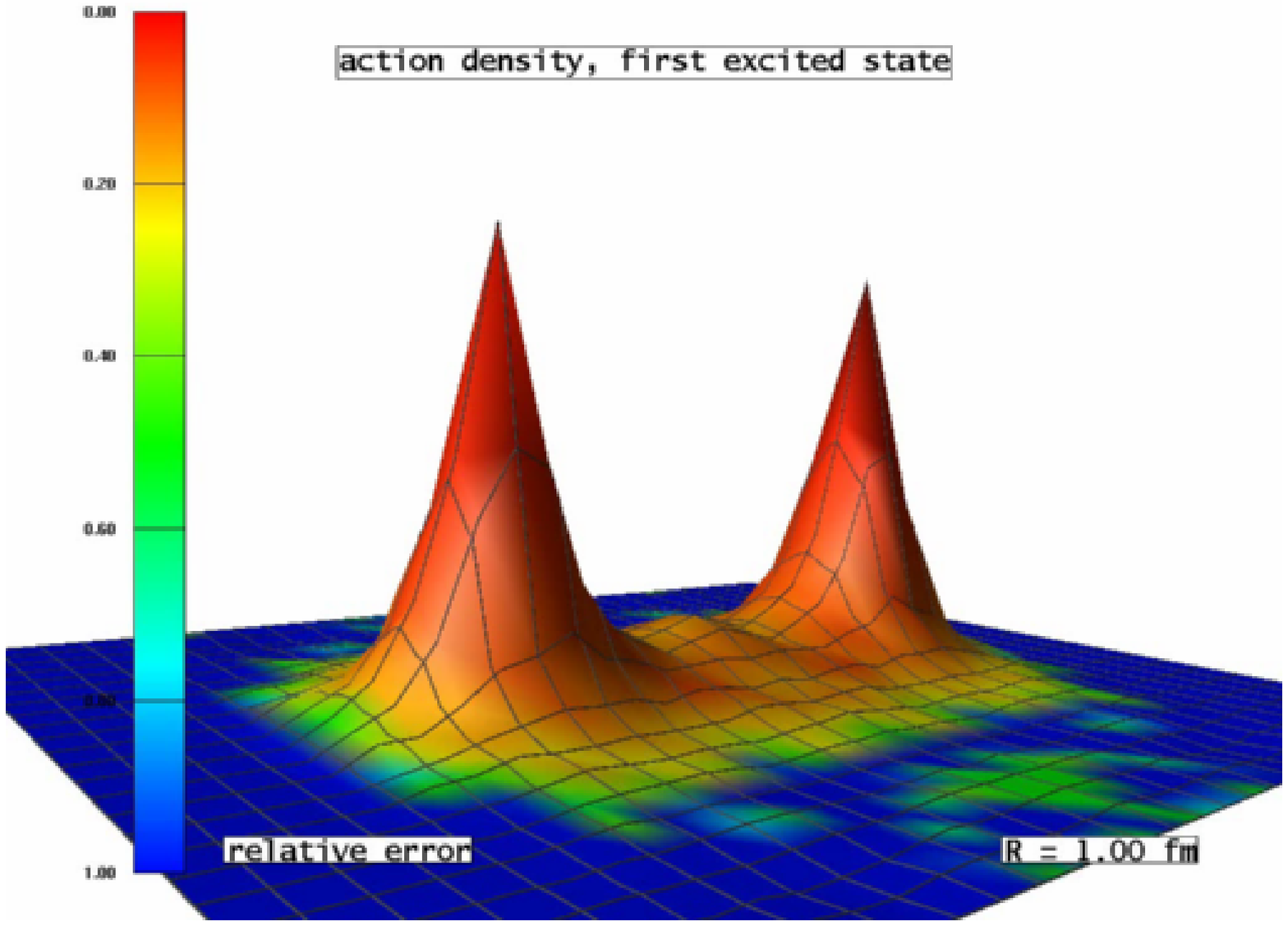,width=0.95\textwidth}}
\end{minipage}
\caption{Action density distribution for the ground state and the first
excitation.}
\label{fig:frame}
\end{figure}
We append two mpeg animations of the action density distribution
as a function
of the distance $1\,\mbox{fm}\leq r\leq 1.5$~fm for the ground state
(string fission) and the excited state (string fusion). The picture frames
inbetween the measurement points of Figure~\ref{fig:pot} are linearly
interpolated. Note that the time is not a linear function of the distance
but dilated within the string breaking region. On a linear
time scale string breaking takes place rather rapidly. The starting frames
are displayed in Fig.~\ref{fig:frame}. The colour encodes the
relative statistical errors and the lattice mesh represents
our spatial resolution $a$. In spite of the fact that the excited
state has (within errors) the energy of two isolated static-light mesons,
a significant string component is already present at
$r<r_c$ which then grows as $r$ is further increased.
This is already obvious from the $\theta(r)$ of Figure~\ref{fig:pot}.
\section{Conclusion}
At any distance the action density looks like a superposition
between (hypothetical) $\overline{Q}Q$
and $B\overline{B}$ distributions
with little interference: light pair creation seems to occur
non-localized and instantaneously. Applying the decay model
of Ref.~\cite{Drummond:1998eh}, where the interaction term is
instantaneous and only depends on the separation $r$, we
undershoot the experimental
$\Upsilon(4S)\rightarrow B\overline{B}$ decay rate by a factor of about
{\em two}. This appears very reasonable, considering the crudeness of
the model and the fact that the gap $\Delta E_c$ will increase
with lighter, more realistic sea quark masses. We are studying the situation
in more detail.

\section*{Acknowledgments}
The computations have been performed on the IBM
Regatta p690+ (Jump) of ZAM at FZ-J\"ulich and on the ALiCE cluster computer
of Wuppertal University. This work is supported by the
EC Hadron Physics I3 Contract RII3-CT-2004-506078,
by the Deutsche Forschungsgemeinschaft and by PPARC.

\end{document}